\documentclass{PoS}

\newcommand{\beq}{\begin{eqnarray}}
\newcommand{\eeq}{\end{eqnarray}}
\def\tr{\mathop{\mathrm{tr}}\nolimits}

\title{Finite-temperature phase transition of $N_{f}=3$ QCD with exact center symmetry
\thanks{Numerical simulation for this study was carried out on Hitachi SR16000 
and IBM System Blue Gene Solution at KEK under its Large-Scale Simulation Program
(No. 14/15-12) and Hitachi SR16000 at YITP in Kyoto University.
E.~I. and T.~I. are supported in part 
by Strategic Programs for Innovative Research
(SPIRE) Field 5.
T.~M. is in part supported by the Japan Society for the 
Promotion of Science (JSPS) Grants Number 26800147. }}

\ShortTitle{Finite-temperature phase transition of $N_{f}=3$ QCD with exact center symmetry}

\author{\speaker{Tatsuhiro Misumi}\\
        Department of Mathematical Science, Akita University \\
Research and Education Center for Natural Sciences, Keio University \\
        E-mail: \email{misumi@phys.akita-u.ac.jp}}

\author{Takumi Iritani\\
        Department of Physics and Astronomy, Stony Brook University (SUNY)\\
        E-mail: \email{takumi.iritani@stonybrook.edu}}

\author{Etsuko Itou\\
        Theory Center, High Energy Accelerator Research Organization (KEK) \\
        E-mail: \email{eitou@post.kek.jp}}

\abstract{
For the $Z_{3}$-symmetric lattice QCD-like theory ($Z_3$-QCD), 
in which $SU(3)$ gauge theory is coupled with
three fundamental Wilson quarks with flavor-dependent twisted boundary conditions,
we calculate the expectation values of Polyakov loop and chiral condensate 
as functions of temperature on $16^3 \times4$ and $20^3 \times 4$ lattices 
with $m_{PS}/m_{V}=0.70$ fixed.
We find the first-order phase transition with respect to the $Z_{3}$ center symmetry, 
where the Polyakov loop exhibits a hysteresis depending on the initial condition of thermalization process.
We also show that the crossover behavior of chiral condensate around the critical temperature of the center transition and the manifestation of flavor symmetry breaking in the high-temperature phase.
}

\FullConference{The 33rd International Symposium on Lattice Field Theory\\
		14 -18 July 2015\\
		Kobe International Conference Center, Kobe, Japan*}

\begin{document}

\section{Introduction}
\label{sec:intro}
For the finite-temperature QCD, 
the interconnection between deconfining and chiral transitions has not been completely understood.
Some of lattice QCD simulations report that the two transition temperatures coincide 
while others do not ~\cite{Kogut:1982rt,Bazavov:2013txa,Fodor:2009ax}.
As well-known, the presence of exact $Z_{3}$ center symmetry
enable us to study the deconfining phase transition 
via the expectation value of the Polyakov loop.
Although the dynamical fundamental quarks explicitly break the center symmetry,
the appropriate boundary conditions for the quarks ($Z_{3}$ t.b.c.) prevents it from breaking: 
Imposing flavor-dependent twisted boundary conditions on the three quarks (shifted by $2\pi/3$) 
in the compact direction leads to the center symmetric $SU(3)$ gauge theory with
three quarks in the fundamental representation on $R^3 \times S^1$ ($Z_{3}$-QCD)
\cite{Kouno:2012zz, Sakai:2012ika, Kouno:2013zr, Kouno:2013mma, Kouno:2015sja}.

In this paper, we study the finite-temperature lattice setup of the $Z_{3}$-QCD model, 
with emphasis on center phase transition and its influence on the chiral transition.
We numerically calculate the expectation value of Polyakov loop 
and chiral condensate on $16^3 \times 4$ and $20^3 \times 4$ lattices with $m_{PS}/m_{V}=0.70$ fixed.
We find out the first-order center phase transition exhibiting the hysteresis 
depending on the thermalization process.
We also show that the crossover behavior of chiral condensate with the hysteresis
and the manifestation of flavor symmetry breaking in the high-temperature phase.
This talk is mainly based on our recent work \cite{Iritani:2015ara}.

\section{Setup}
\label{sec:setup}
The continuum setup of $Z_{3}$-QCD model is investigated in
\cite{Kouno:2012zz, Sakai:2012ika, Kouno:2013zr, Kouno:2013mma, Kouno:2015sja},
and we just show the lattice setup of the model we have used in the lattice simulation.
Our setup is based on the Iwasaki gauge action with naive Wilson fermions.
The lattice action is given by
\beq
S&=&S_g + S_f, \nonumber\\
S_g &=& \beta \sum_x \left( c_0 \sum^4_{\mu < \nu; \mu,\nu=1} W_{\mu \nu}^{1 \times 1} (x) +c_1 \sum^4_{\mu \ne \nu;\mu,\nu=1} W^{1 \times 2}_{\mu \nu} (x)   \right),\\
S_f&=&\sum_{f=u,d,s} \sum_{x,y} \bar{\psi}^f_x M_{x,y} \psi^f_y,\label{eq:fermi-action}
\eeq
with $\beta=6/g^2$. Here $g$ is a bare gauge coupling constant and the parameters for the Iwasaki gauge action are $c_1=-0.331$ and $c_0=1-8c_1$.
$W^{1\times 1}$ and $W^{1 \times 2}$ denote the plaquette and the rectangular, respectively. 
$M_{x,y}$ is given by
\beq
M_{x,y}&=&\delta_{x,y} -\kappa \sum_{\mu=1}^4 \left\{ (1-\gamma_\mu) U_{x,\mu}\delta_{x+\hat{\mu},y} + (1+\gamma_\mu) U^\dag_{y,\mu} \delta_{x,y+\hat{\mu}}  \right\},
\label{eq:def-Dirac}
\eeq
where $\kappa$ is the hopping parameter.

To implement the $Z_{3}$ twisted boundary condition on the lattice, 
we introduce the following twisted boundary conditions for the link variable in the fermion action 
Eq.~(\ref{eq:fermi-action}):
\beq
U_4 (\vec{x}, \tau=N_\tau) =& -  U_4 (\vec{x}, \tau=0) & \mbox{ for $u$-flavor }, \nonumber \\
U_4 (\vec{x}, \tau=N_\tau) =& - e^{2 \pi i/3} U_4 (\vec{x}, \tau=0) &\mbox{ for $d$-flavor}, \nonumber \\
U_4 (\vec{x}, \tau=N_\tau) =& -  e^{4 \pi i / 3}U_4 (\vec{x}, \tau=0) & \mbox{ for $s$-flavor}.
\eeq
These conditions are equivalent to the flavor-dependent imaginary chemical potential 
for the standard finite-temperature QCD; $\mu_I=0, 2\pi /3$ and $4 \pi/3$, respectively.
As shown in \cite{Kouno:2012zz, Sakai:2012ika, Kouno:2013zr, Kouno:2013mma, Kouno:2015sja}, 
the flavor-chiral symmetry $SU(3)_{L}\times SU(3)_{R}$
is broken to its Cartan subgroup $U(1)^{2}_{L}\times U(1)^{2}_{R}$ 
due to the $Z_{3}$ twisted boundary condition.
In our simulation, we use the Rational Hybrid Monte Carlo (RHMC) algorithm 
to calculate the fermion determinant for each flavor.

Before looking into the finite-temperature simulation, we show the way how to fix the parameters.
We perform the zero-temperature simulation on $16^4$ lattices to obtain the line of constant-physics.
We carry out the simulations for several values of the hopping parameter $\kappa$ with each $\beta$ 
and measure the pseudo-scalar mass ($m_{PS}$) and vector meson mass ($m_V$) for 
each flavor. We generate $2,000$ -- $3,000$ trajectories, and measure the correlator of these hadronic states every $10$ Monte Carlo trajectories, where the estimated autocorrelation length is about $100$ trajectories.

\begin{table}[h]
\begin{center}
\begin{tabular}{|c|c|c|c|c|c|c|c|c|c|c|c|c|c|c|}
\hline
$\beta$    &$1.30$      & $1.40$       & $1.50$         & $1.60$  & $1.70$     & $1.80$ & $1.90$ & $2.00$ & $2.10$ & $2.20$ \\  
\hline
$\kappa$ & $0.2019$  &  $0.1975$ & $0.1921$   & $0.1861$&  $0.1793$ & $0.1725$ & $0.1663$ &  $0.1611$ & $0.1571$ & $0.1539$ \\
\hline
\end{tabular}
\caption{Simulation parameters corresponding 
to $m_{PS}/m_{V}=0.70$: $\beta$ and $\kappa$.
}  \label{table:sim-parameter}
\end{center}
\end{table}

We fix the ratio between $m_{PS}$ and $m_V$ as $m_{PS}/m_{V}=0.70$,
and tune the value of the hopping parameter for each $\beta$ shown in Table~\ref{table:sim-parameter}.
We use these parameter sets in Table~\ref{table:sim-parameter} in 
the finite temperature simulation on $16^3 \times 4$ and $20^3\times 4 $.
The number of trajectories for the finite temperature simulation is $500$ -- $5,000$.
We measure the Polyakov loop in the temporal direction for every Monte Carlo trajectory 
and the chiral condensate every $10$ trajectories. 
We note that $m_{PS}$ and $m_{V}$ are the same as those 
of the standard three-flavor QCD at zero temperature, which indicates 
that the breaking of $SU(3)_{L}\times SU(3)_{R}$ flavor-chiral symmetry due to
the $Z_{3}$ twisted boundary condition is not manifest at zero temperature.

\section{Results}
\label{sec:results}

\subsection{Polyakov loop}
\begin{figure}
\centering
\includegraphics[width=0.4\textwidth]{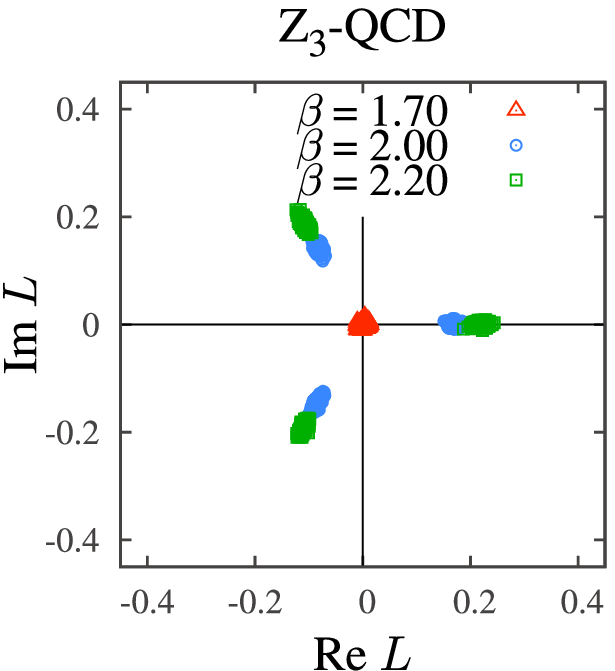} 
\hspace{5mm}
\includegraphics[width=0.4\textwidth]{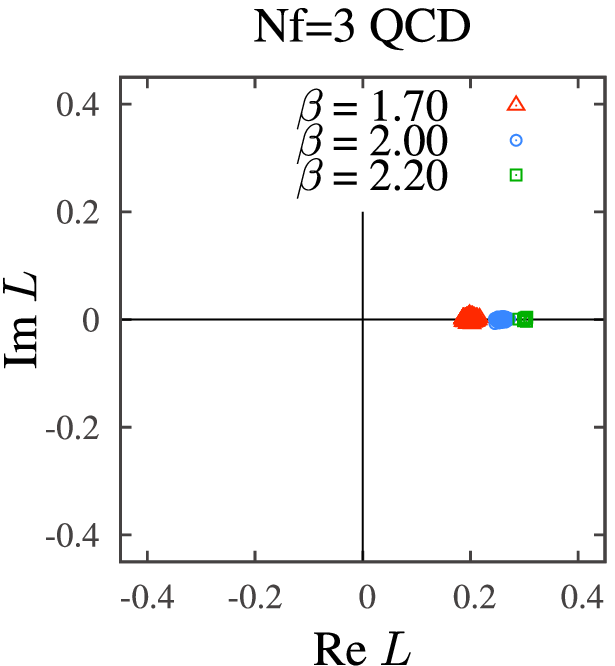}
\caption{Polyakov loop distribution plot in $Z_{3}$-QCD (left) and the standard three-flavor QCD (right).
Both are based on $16^3 \times 4$ lattice for $\beta=1.70,2.00,2.20$ with the same values of $\kappa$.
}
\label{fig: Pdis1}
\end{figure}

We first show the distribution plot of the following Polyakov loop ($L$)
which exhibits exact $Z_{3}$ center symmetry in the present model.
\beq
L= \frac{1}{V} \sum_{\vec{x}} \frac{1}{3} \tr \left[ \prod_{i=1}^{N_\tau} U_\tau (\vec{x},i) \right],
\eeq
where $V$ denotes the spatial volume in a lattice unit.
The left panel of Fig.~\ref{fig: Pdis1} for $Z_{3}$-QCD shows that 
the Polyakov loops are distributed around the origin
in the low $\beta$ regime while three vacua exist in the high $\beta$ regime.
On the other hand, in the right panel of Fig.~\ref{fig: Pdis1}, those in the standard three-flavor 
$SU(3)$ gauge theory shows explicit breaking of $Z_{3}$ center symmetry.
Our plot indicates that the $Z_{3}$-QCD model possesses exact $Z_{3}$ center symmetry,
then the symmetry is spontaneously broken in the high-temperature phase.

\begin{figure}
\centering
\includegraphics[width=0.45\textwidth]{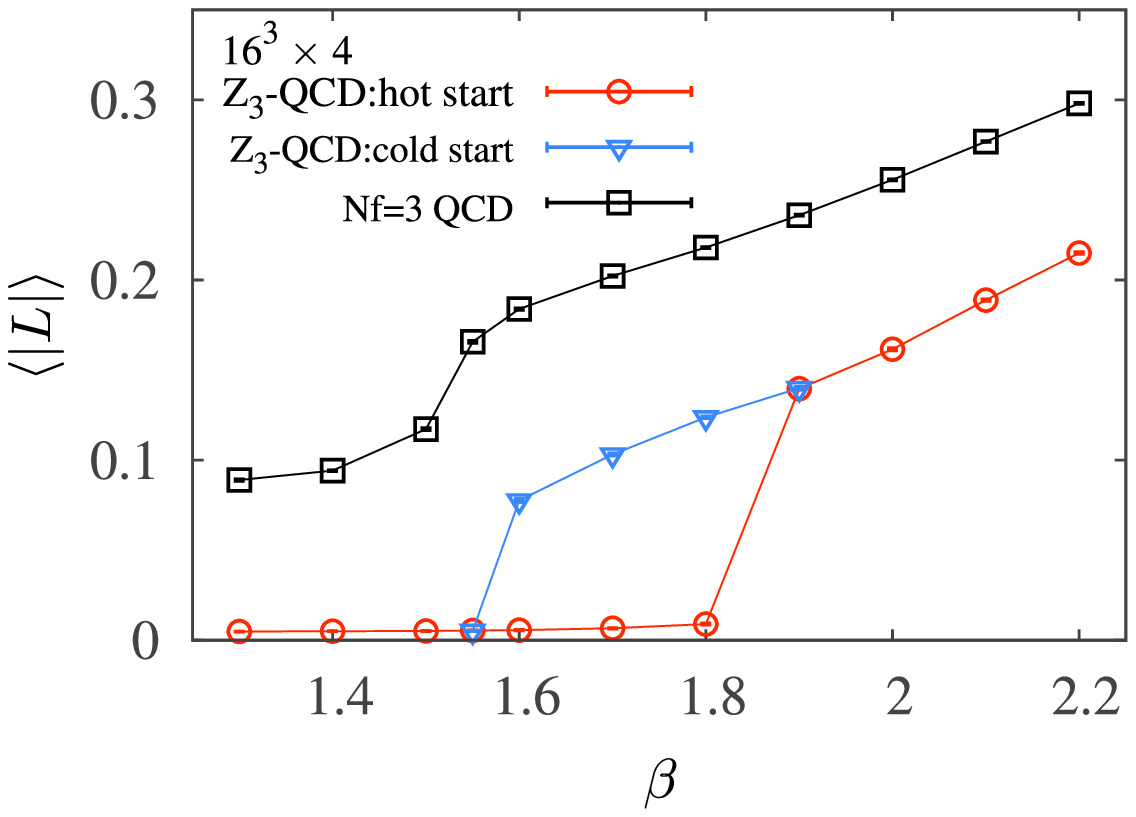} 
\hspace{5mm}
\includegraphics[width=0.45\textwidth]{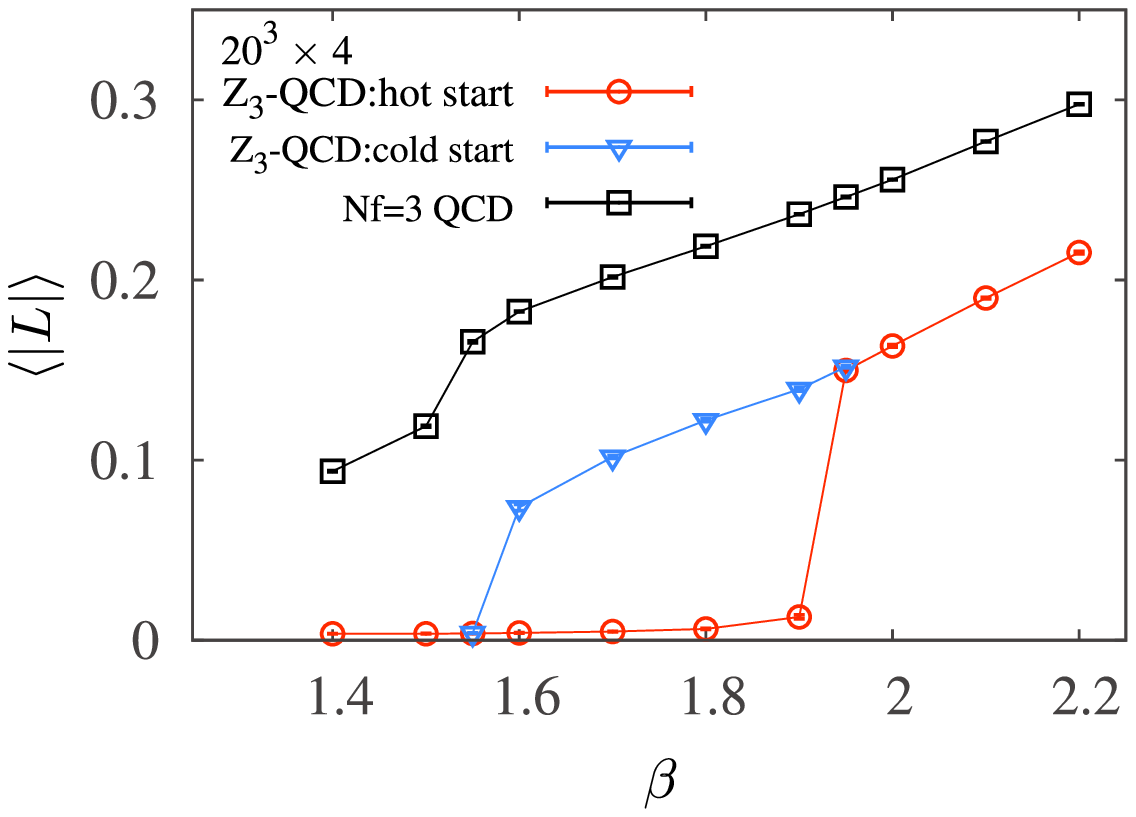} 
\caption{ 
$\beta$ dependence of the magnitude of Polyakov loop ($\langle |L| \rangle$)
for the $Z_{3}$-QCD and standard three-flavor QCD on $16^3 \times 4$ (left) and $20^3\times 4$ (right) lattices.
For the $Z_3$-QCD model, the data of $\langle |L| \rangle$ started with the cold start (triangle blue symbols) have a clear jump from zero to non-zero values around the region $1.55 \le \beta \le 1.60$ in both panels,
while the jump occurs in $1.80 \le \beta \le 1.90$ (left) and $1.90 \le \beta \le 1.95$ (right) for the data generated by the hot start (circle red symbols).
A clear hysteresis stands between these two jumps in $Z_3$-QCD model.
On the other hand, the data of the standard three-flavor QCD (square black symbols) do not show such a jump nor a hysteresis.
}
\label{fig:Pb}
\end{figure}

We now investigate dependence of the Polyakov loop on temperature by varying $\beta$
with $m_{\rm PS}/m_{V}=0.70$ fixed (see Table~\ref{table:sim-parameter}).
We generate configurations based on the two types of initial conditions; cold start and hot start.
In both panels of Fig.~\ref{fig:Pb}, the triangle (blue) symbol denotes the data started with cold start. The corresponding initial configuration lives in the ordered phase, and we set all initial link variables to unity. 
On the other hand, the circle (red) symbol denotes the ones started with hot start. The corresponding configuration is in the disordered phase, and the initial link variable is a random number. 
The square (black) symbol shows the result of the standard finite-temperature three-flavor QCD
with the same parameter set as $Z_{3}$-QCD simulations. 

Our results are summarized as follows:
In the low-temperature regime, the magnitude of Polyakov loop is exactly zero for the $Z_3$-QCD,
and it gets nonzero in the high-temperature regime.
This behavior of Polyakov loop is clearly different from that of the usual three-flavor QCD.
Moreover, for the $Z_3$-QCD model, we find the hysteresis of Polyakov loop 
in the range of $1.55 < \beta < 1.90$
depending on the initial conditions (cold or hot) while there is no hysteresis in the standard three-flavor QCD.
Such a hysteresis is one of the signals of the first-order phase transition.
These results indicate that the $Z_3$-QCD model undergoes first-order phase transition, 
where the $Z_{3}$ center symmetry is spontaneously broken.
Due to the limitation of space, we omit to discuss our result on 
the Polyakov loop susceptibility \cite{Iritani:2015ara} defined by
\beq
\chi_{\langle | L | \rangle} \equiv V  \left[ \langle |L|^2 \rangle - \langle |L| \rangle^2 \right],
\eeq
which helps determine the transition temperature.
The precise determination of the critical temperature in the large volume and continuum limits 
remains as a future work.

\subsection{Chiral condensates}
We here investigate the chiral transition in the finite-temperature $Z_3$-QCD model, 
which is characterized by the chiral condensate.
Due to the $Z_{3}$ twisted boundary condition, 
the pattern of symmetry breaking is expected to 
be $U(1)^{2}_L \times U(1)^{2}_R \to U(1)^{2}_{\bar{V}}$ in the present model.
On the other hand, the result based on the effective chiral model implies 
that the flavor symmetry breaking due to the $Z_{3}$ twisted boundary condition 
becomes manifest only in the high-temperature 
phase~\cite{Kouno:2012zz, Sakai:2012ika, Kouno:2013zr, Kouno:2013mma, Kouno:2015sja} 
while it does not in the low-temperature phase.
To elucidate this conjecture, we calculate the chiral condensate for each flavor.
We note that, since our simulation is based on the heavy mass $m_{\rm PS}/m_{V}=0.70$, 
the decrease of chiral condensate just indicates 
effective and approximate restoration of chiral symmetry.

We consider the following flavor-diagonal expectation value 
of the subtracted chiral condensate \cite{Giusti:1998wy, Umeda:2012nn, Hayakawa:2013maa} for each flavor
\cite{Bochicchio:1985xa, Itoh:1986gy, Aoki:1997fm}
\beq
 \langle \bar{\psi}^f \psi^f\rangle_{\mathrm{subt.}} &=&  (2m_{\rm PCAC})(2\kappa)^2 \sum_{x} \langle P(\vec{x},t) P^\dag(\vec{0},0) \rangle.   \label{eq:def-chiral}  
\eeq
Here, $m_{PCAC}$ is partially conserved axial current (PCAC) mass, and $P$ denotes the pseudo scalar state defined as $P \equiv \bar{\psi}^f \gamma_5 \psi^f$ for the flavor ($f=u,d,s$) in the left hand side.
The PCAC mass is defined via axial Ward identity;
\beq
2m_{PCAC} =\frac{ \sum_{\vec{x}} \partial_4 \langle A_4 (\vec{x},t) P^\dag (\vec{0},0) \rangle }{ \sum_{\vec{x}} \langle P(\vec{x},t) P^\dag(\vec{0},0) \rangle },
\eeq
where $A_\mu$ corresponds to the axial vector current defined by 
$A_\mu =\bar{\psi}^f \gamma_5 \gamma_\mu \psi^f$.

\begin{figure}
\centering
\includegraphics[width=0.45\textwidth]{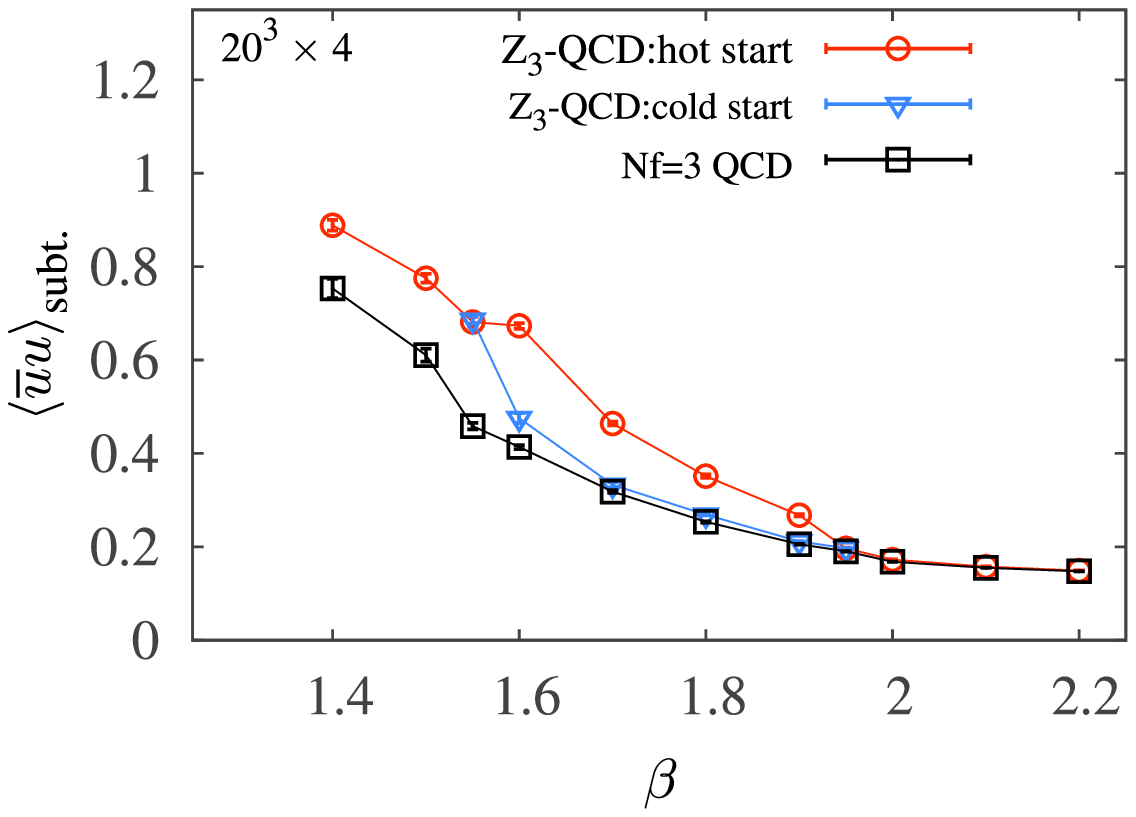} 
\hspace{5mm}
\includegraphics[width=0.45\textwidth]{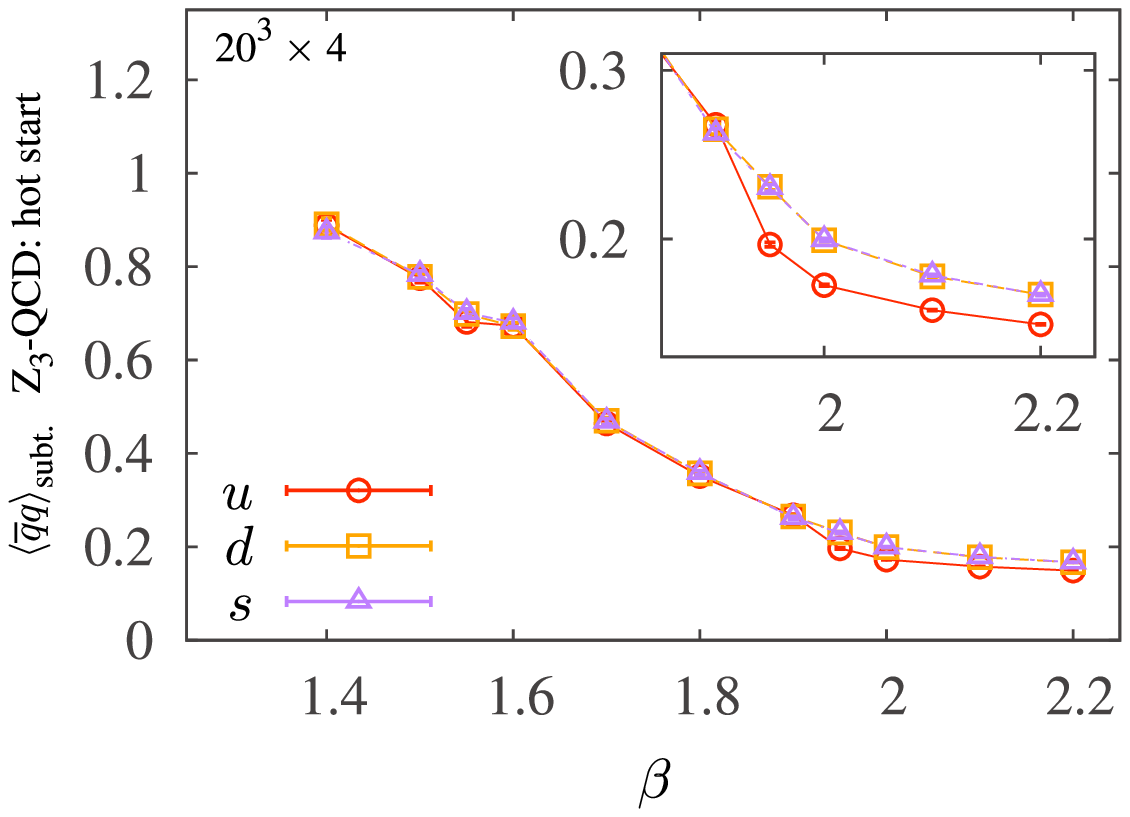} 
\caption{ 
{\it Left panel}: $\beta$ dependence of the expectation values of subtracted chiral condensates 
$\langle \bar{u} u \rangle_{\mbox{subt.}}$ for $Z_{3}$-QCD and three-flavor QCD on $20^3 \times 4$ lattices.
Circle (red), triangle (blue) and square (black) symbols stand for those data associated with the hot start and cold start in $Z_3$-QCD and the three-flavor QCD, respectively.
{\it Right panel}: $\langle\bar{q}q \rangle$ for each flavor in $Z_{3}$-QCD model on $20^3 \times 4$.
Circle (red), square (orange) and triangle (violet) symbols denote $u$-, $d$- and $s$-flavor generated 
with the hot start, respectively.}
\label{fig:chiral-cond}
\end{figure}

Our results on chiral condensate are summarized as follows: 
In Fig.~\ref{fig:chiral-cond}(left) we show $\beta$ dependence of chiral condensate for $u$-flavor 
in the $Z_{3}$- and three-flavor QCD.
Circle (red), triangle (blue) and square (black) symbols stand for the data associated with the hot start and cold start in $Z_3$-QCD model and the standard three-flavor QCD, respectively.
For all the three cases, the chiral condensate gradually decreases as $\beta$ increases.
However, $\beta$ dependence of chiral condensates in $Z_3$-QCD
exhibits a clear hysteresis depending on the initial conditions as with that of the Polyakov loop.
It implies that effective restoration of the chiral symmetry is progressing from $\beta = 1.55$ to 
$\beta=1.95$ for $Z_3$-QCD.
The decreasing rate of chiral condensate in $Z_3$-QCD model seems 
larger than the one in the standard three-flavor QCD.
As for the chiral crossover temperature,
it is not clear whether or not the chiral transition temperatures 
in $Z_{3}$-QCD and three-flavor QCD coincide.
(See \cite{Borsanyi:2010bp, Bazavov:2011nk, Bhattacharya:2014ara} 
on the determination of the chiral transition temperature.)
According to the arguments in Refs.~\cite{Roberge:1986mm, D'Elia:2002gd},
the $2\pi/3$ periodicity of partition function in the imaginary chemical potential
could lead to coincidence of the chiral phase transition temperature in the two theories.
On the other hand, since the flavor-chiral symmetry in $Z_{3}$-QCD model in the chiral limit 
is broken to $U(1)^{2}_{L} \times U(1)^{2}_{R}$ due to the twisted boundary condition,
we may have the smaller number of Nambu-Goldstone modes than the usual three-flavor QCD,
which lifts the phase transition temperature in general.
We need further investigation on the topic. (See \cite{Kratochvila:2006jx} for the similar situation.)

In Fig.~\ref{fig:chiral-cond}(right), we show the expectation values of chiral condensates for each flavor.
Circle (red), square (orange) and triangle (violet) symbols denote $u$-, $d$- and $s$-flavor generated 
with the hot start, respectively.
The figure shows that the three components of chiral condensates are 
degenerate in the low-temperature phase, while in the high-temperature phase 
there is clear flavor symmetry breaking.
It means that the $Z_3$ center of $SU(3)$ flavor symmetry is effectively intact
in the low-temperature phase, while the breaking of this symmetry gets manifest in the high-temperature phase.
We need further study to verify the conjecture of the chiral model, 
which states that the $SU(3)$ flavor symmetry is not influenced by the twisted boundary condition 
in the low-temperature phase
\cite{Kouno:2012zz, Sakai:2012ika, Kouno:2013zr, Kouno:2013mma, Kouno:2015sja}. 
What we can argue from our results is that the effective preservation of the $Z_{3}$ flavor symmetry 
in the low-temperature phase is consistent with the conjecture.

\section{Summary and Discussion}
\label{sec:summary}
We investigate the finite-temperature $Z_{3}$-QCD model on the lattice, 
with emphasis on center phase transition and its influence on the chiral transition.
Temperature dependence of Polyakov loop 
and chiral condensate are calculated on $16^3 \times 4$ and $20^3\times 4$ 
lattices with $m_{PS}/m_{V}=0.70$ fixed.
We find out the signal of first-order center phase transition,
which is characterized by the hysteresis depending on the initial conditions.
The chiral crossover transition takes place around the center critical temperature, 
where the chiral condensate exhibits a hysteresis in the same range as that of the center phase transition.
We find that the decrease of chiral condensate in $Z_{3}$-QCD is more rapid 
than that of the standard three-flavor QCD while temperatures 
of the two crossover transitions are almost the same.
Our results also support the manifestation of flavor symmetry breaking 
due to the $Z_{3}$ twisted boundary condition
in the high-temperature phase, which has been conjectured in
\cite{Kouno:2012zz, Sakai:2012ika, Kouno:2013zr, Kouno:2013mma, Kouno:2015sja}.

\end{document}